\documentclass[10pt,aps,prd,nofootinbib,superscriptaddress,twocolumn,preprintnumbers,balancelastpage]{revtex4}
\usepackage{amssymb,amsmath,latexsym,graphics, graphicx,epsfig,multirow,comment,appendix, cancel} 
\usepackage{setspace}

\usepackage{color}
\definecolor{Mahogany}{rgb}{0.62,0.24,0.15}
\definecolor{colorLink}{rgb}{0.7,0,0}
\definecolor{colorCite}{rgb}{0,.7,0}
\definecolor{colorURL}{rgb}{0,0,0.7}
\usepackage[colorlinks=true,linktocpage=true,linkcolor=black,citecolor=colorCite
,urlcolor=colorURL]{hyperref}

\def\be{\begin{equation}}
\def\ee{\end{equation}}
\newcommand{\beq}{\begin{equation}}
\newcommand{\eeq}{\end{equation}}
\def\bea{\begin{eqnarray}}
\def\eea{\end{eqnarray}}
\newcommand{\eref}[1]{Eq.~(\ref{#1})}

\newcommand{\TeV}{{\text{ TeV}}}

\newcommand{\ME}{\diagup \!\!\!\!\! E}
\newcommand{\vME}{\diagup \!\!\!\!\! \vec{E}}

\usepackage{color}

\newcommand{\SM}{{\rm SM}}
\newcommand{\D}{{d}}

\begin{document}
\title{Semi-visible Jets:  Dark Matter Undercover at the LHC}
\author{Timothy Cohen}
\affiliation{Department of Physics, Princeton University, Princeton, NJ 08544}
\affiliation{School of Natural Sciences, Institute for Advanced Study, Princeton, NJ, 08540}
\affiliation{Institute of Theoretical Science, University of Oregon, Eugene, OR, 97403}
\author{Mariangela Lisanti}
\affiliation{Department of Physics, Princeton University, Princeton, NJ 08544}
\author{Hou Keong Lou}
\affiliation{Department of Physics, Princeton University, Princeton, NJ 08544}

\begin{abstract}
The dark matter may be a composite particle that is accessible via a weakly coupled portal.  If these hidden-sector states are produced at the Large Hadron Collider (LHC), they would undergo a QCD-like shower.  This would result in a spray of stable invisible dark matter along with unstable states that decay back to the Standard Model.  Such ``semi-visible" jets arise, for example, when their production and decay are driven by a leptophobic $Z'$ resonance; the resulting signature is characterized by significant missing energy aligned along the direction of one of the jets.  These events are vetoed by the current suite of searches employed by the LHC, resulting in low acceptance.  This Letter will demonstrate that the transverse mass---computed using the final-state jets and the missing energy---provides a powerful discriminator between the signal and the QCD background.  Assuming that the $Z'$ couples to the Standard Model quarks with the same strength as the $Z^0$, the proposed search can discover (exclude) $Z'$ masses up to 2.5~TeV (3.5~TeV) with 100~fb$^{-1}$ of 14 TeV data at the LHC.
\end{abstract}
\maketitle

The existence of dark matter provides one of the strongest motivations for physics beyond the Standard Model, and its discovery is one of the core missions for the Large Hadron Collider (LHC) program.  Under the assumption that the dark-matter particle is neutral and stable, it escapes the detector and manifests as missing transverse energy ($\ME_T$).  The LHC collaborations have developed a comprehensive search strategy to look for signals with significant $\ME_T$, accompanied by jets and/or leptons (see,~\emph{e.g.}~\cite{Mitsou:2013rwa} for a review).  These searches are typically cast in terms of a Simplified Model~\cite{Alves:2011wf} for supersymmetry or an effective theory of dark-matter interactions~\cite{Goodman:2010yf, Fox:2011pm}.  Yet if one relaxes the assumption that the dark sector is weakly coupled, a new class of dark-matter signatures emerge that evade this entire suite of analyses.  Namely, it is possible that the dark matter has been lurking undercover within hadronic jets.  The purpose of this Letter is to propose a straightforward discovery strategy for these ``semi-visible" jets.

Semi-visible jets may occur if the dark matter is the stable (or meta-stable) remnant of a more complicated dark sector.  
  The dynamics of non-trivial dark sectors have been explored in many contexts, \emph{e.g.}~\cite{Chacko:2005pe, Schabinger:2005ei, Strassler:2006im, Burdman:2006tz, Kang:2008ea, MarchRussell:2008yu, ArkaniHamed:2008qn, Kaplan:2009ag, Kribs:2009fy, McDonald:2001vt, Barger:2007im,Porto:2007ed, Bertolami:2007wb, Hochberg:2014dra, Alves:2013tqa}.  In these models, the dark sector contains a dark-matter candidate(s) and possibly new force carriers and/or matter fields.  Note that we are agnostic about how much of the cosmological relic density is accounted for by this dark-matter candidate.  Messenger states that couple the dark sector to the Standard Model (SM) can exist.  If the messenger is accessible at colliders, dark-sector states can be produced, leading to unique signatures such as large particle multiplicities, displaced vertices, multiple resonances, and lepton or photon jets~\cite{Strassler:2006qa, Strassler:2006ri, Han:2007ae, Strassler:2008fv, ArkaniHamed:2008qp,  Kilic:2009mi, Baumgart:2009tn, Chang:2009sv, Carloni:2010tw, Carloni:2011kk, Chan:2011aa, Chatrchyan:2013xva, Aad:2014nra, Schwaller:2015gea}.  
  
Another possibility is that the final state resulting from strongly coupled hidden sectors may contain a new type of jet object---a semi-visible jet.  
In this case, the dark matter is produced in a QCD-like parton shower along with other light degrees of freedom that decay hadronically.  The result is a multijet+$\ME_T$ signature where one of the jets is closely aligned with the $\vME_T$.  A cornerstone of the standard multijet+$\ME_T$ searches is to require a minimum angular separation between the jets and $\vME_T$ to remove QCD background contamination arising from jet-energy mis-measurement~\cite{Chatrchyan:2014lfa, Aad:2014wea}.  This implies that events containing semi-visible jets have a low acceptance for the currently implemented suite of searches.

\begin{figure*}[t!]
%width=0.48
\includegraphics[width=0.45 \textwidth, trim = 0mm 12mm 0mm 3mm]{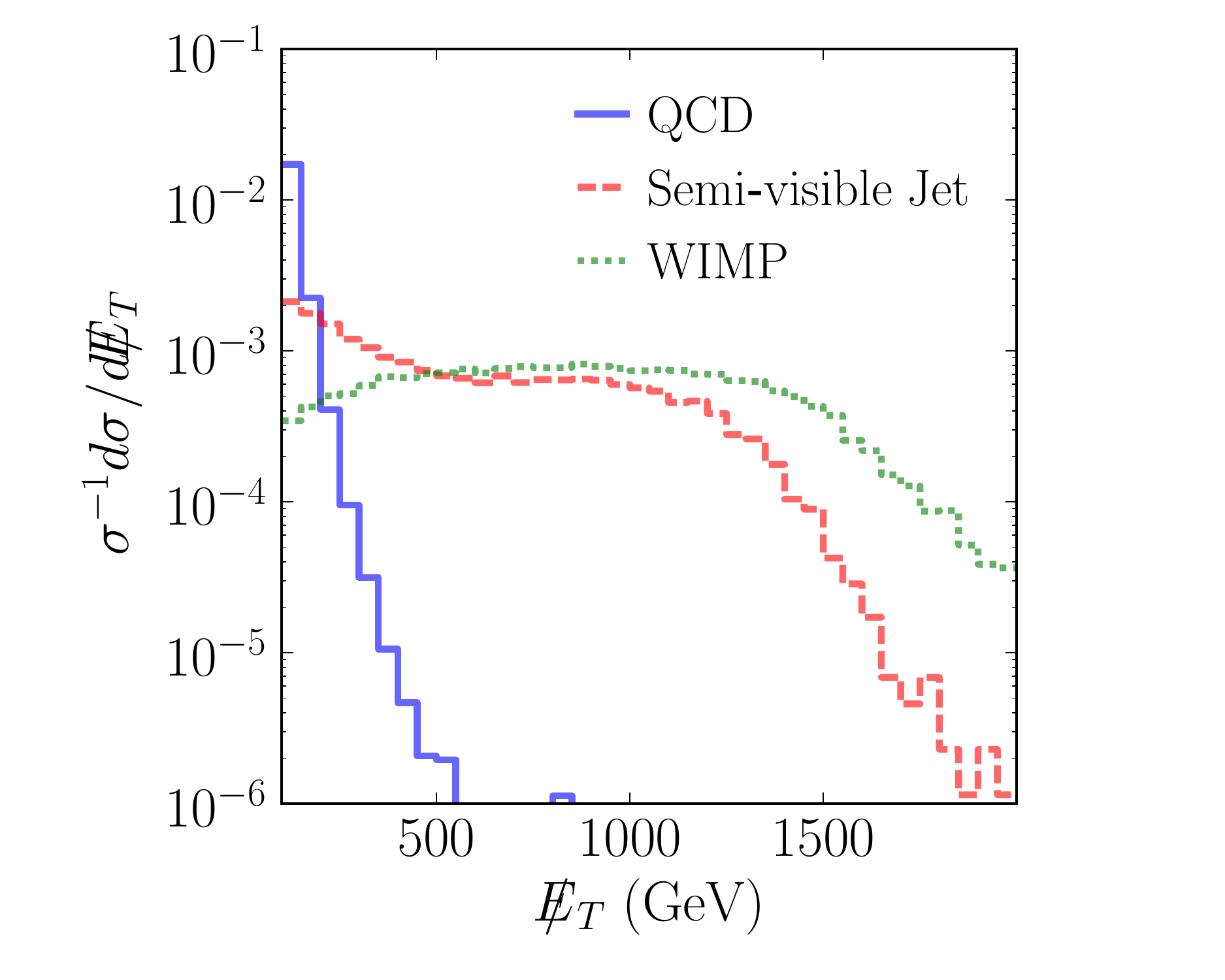}
\includegraphics[width= 0.45 \textwidth, trim = 0mm 12mm 0mm 3mm]{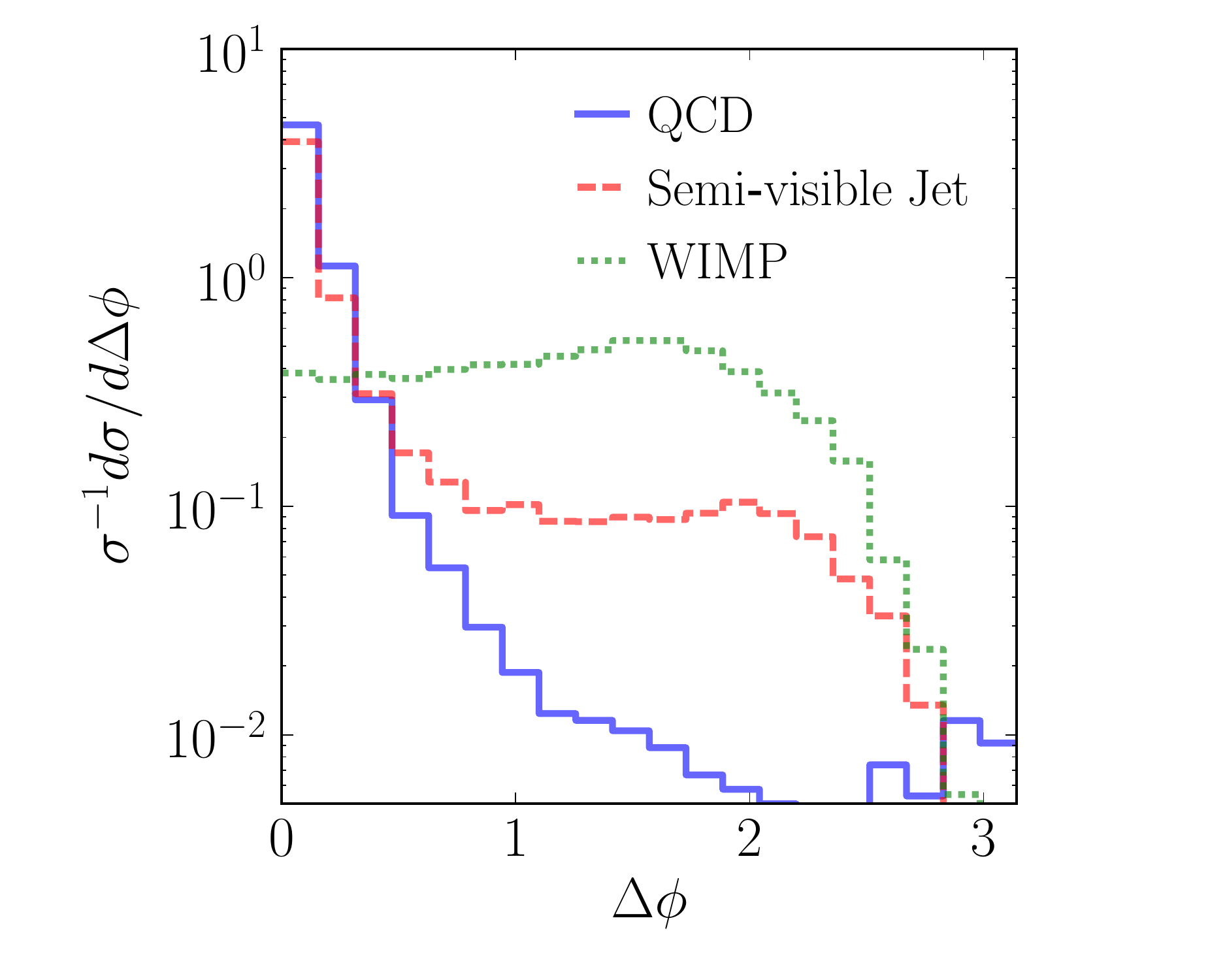}
\caption{\emph{(left)}  The distribution of transverse missing energy $\ME_{T}$ for the QCD background (solid blue), as well as the semi-visible jet (dashed red) and WIMP (dotted green) examples.  \emph{(right)} The distribution of $\Delta \phi~\!\equiv~\!\text{min}\left\{\Delta\phi_{j_1 \ME_T}, \Delta\phi_{j_2 \ME_T}\right\}$, where $j_{1,2}$ are the two hardest jets. } %Typical LHC searches require $\Delta \phi \gtrsim 0.4$~\cite{Chatrchyan:2014lfa, Aad:2014wea}.}
\label{fig:dphi}
\vspace{-0.15in}
\end{figure*}

To further illustrate this point, Fig.~\ref{fig:dphi} compares selected observables for QCD with those for example weakly coupled and strongly coupled dark-matter models.  The weakly coupled model is derived from supersymmetric theories and results from pair production of 1.5 TeV scalar quark partners.  Each squark decays to a jet and 1 GeV neutral dark-matter particle.  The signal from the strongly coupled model, which will be described more fully later, comes from the production of a 3 TeV resonance which then decays to a pair of dark-sector particles that subsequently shower and hadronize, yielding semi-visible jets. Both these examples yield topologies with jets and missing energy.  As the left panel shows, the weakly coupled (labeled WIMP) and strongly coupled (labeled semi-visible jet) dark-sector models produce considerable $\ME_T$, with tails that extend beyond the QCD distribution.  However, $\Delta\phi\equiv\text{min}\left\{\Delta\phi_{j_1 \ME_T}, \Delta\phi_{j_2 \ME_T}\right\}$, where $j_{1,2}$ are the two hardest jets, is different between these models, as illustrated in the right panel.  The $\Delta \phi$ distribution falls relatively steeply for the strongly coupled case, while it remains relatively flat for the weakly coupled scenario.  Typical LHC searches require $\Delta \phi \gtrsim 0.4$~\cite{Chatrchyan:2014lfa, Aad:2014wea}.  After requiring $\ME_{T} > 500$~GeV and $\Delta \phi > 0.4 $, the acceptance of the WIMP (semi-visible) example is $\sim$70\% (7\%).  We also verified that the $\alpha_T$ underperforms on semi-visible jet signals; a standard cut of $\alpha_T > 0.55$ gives $\sim 20\%$ (3\%) efficiency for the WIMP (semi-visible) scenario \cite{Khachatryan:2015pwa}.  Razor analyses~\cite{CMS:2015fla} may prove useful, but require optimization for semi-visible jets---see Supplementary Material.  

To regain sensitivity to final states containing semi-visible jets, the cut on the angular separation $\Delta \phi$ must be removed.  This comes at the expense of an unsuppressed QCD multijet background, which must be eliminated using other techniques.  In this Letter, we focus on the case where the dark sector is accessed via a heavy resonance.  In such scenarios, one can take advantage of structure in the transverse mass---calculated using the final-state jets and $\ME_T$---to distinguish the signal from QCD.  The strategy employed is similar to others proposed for semi-visible Higgs decays \cite{Englert:2012wf}.  

We now introduce an example Hidden Valley~\cite{Strassler:2006im} model that will enable us to analyze the LHC sensitivity for semi-visible jets.  %We now introduce example messenger and dark-sector Lagrangians which will enable us to analyze the LHC sensitivity for semi-visible jets.  In particular, the example studied below first appeared in the context of ``Hidden Valleys"~\cite{Strassler:2006im}.  
This model is presented for illustration and concreteness; semi-visible jets will be among the LHC signatures for a vast class of dark-sector theories.  The messenger sector is described by a simple phenomenological model for a TeV-scale $U(1)'$ gauge boson.  The new leptophobic $Z'$ gauge boson couples to the SM baryon current $J_{\SM}^\mu$:
\begin{equation}
\mathcal{L} \supset - \frac{1}{4}\,Z'^{\mu\nu}\,Z'_{\mu\nu} - \frac{1}{2} \,M_{Z'}^2 \,Z'_{\mu}\,Z'^{\mu}  - g_{Z'}^{\SM} \,Z'_{\mu}\, J_{\SM}^\mu .
\label{eq:portal}
\end{equation}
Note that the $Z'$ is treated as a Stueckelberg field---the Higgs sector has been neglected as it is not relevant for the LHC phenomenology discussed below; the additional matter needed to render the $U(1)$ of baryon number anomaly free is also ignored.

The dark sector is an $SU(2)_\D$ gauge theory with coupling $\alpha_\D$ and two fermionic quark flavors $\chi_i = \chi_{1,2}$ with masses $M_{i}$.  The dark quark coupling to the $Z'$ is $g^{\D}_{Z'}$.  In general, the couplings $g^{\D}_{Z'}$ and $g_{Z'}^{\SM}$ do not have to be comparable; we focus on the case where $g^{\D}_{Z'}$ is large so that the $Z'$ decays frequently to the dark sector.   

\begin{figure*}[t!]
\centering
%width = 1.0
\includegraphics[width=1.0\textwidth, trim = 1cm 10mm 0mm 10mm]{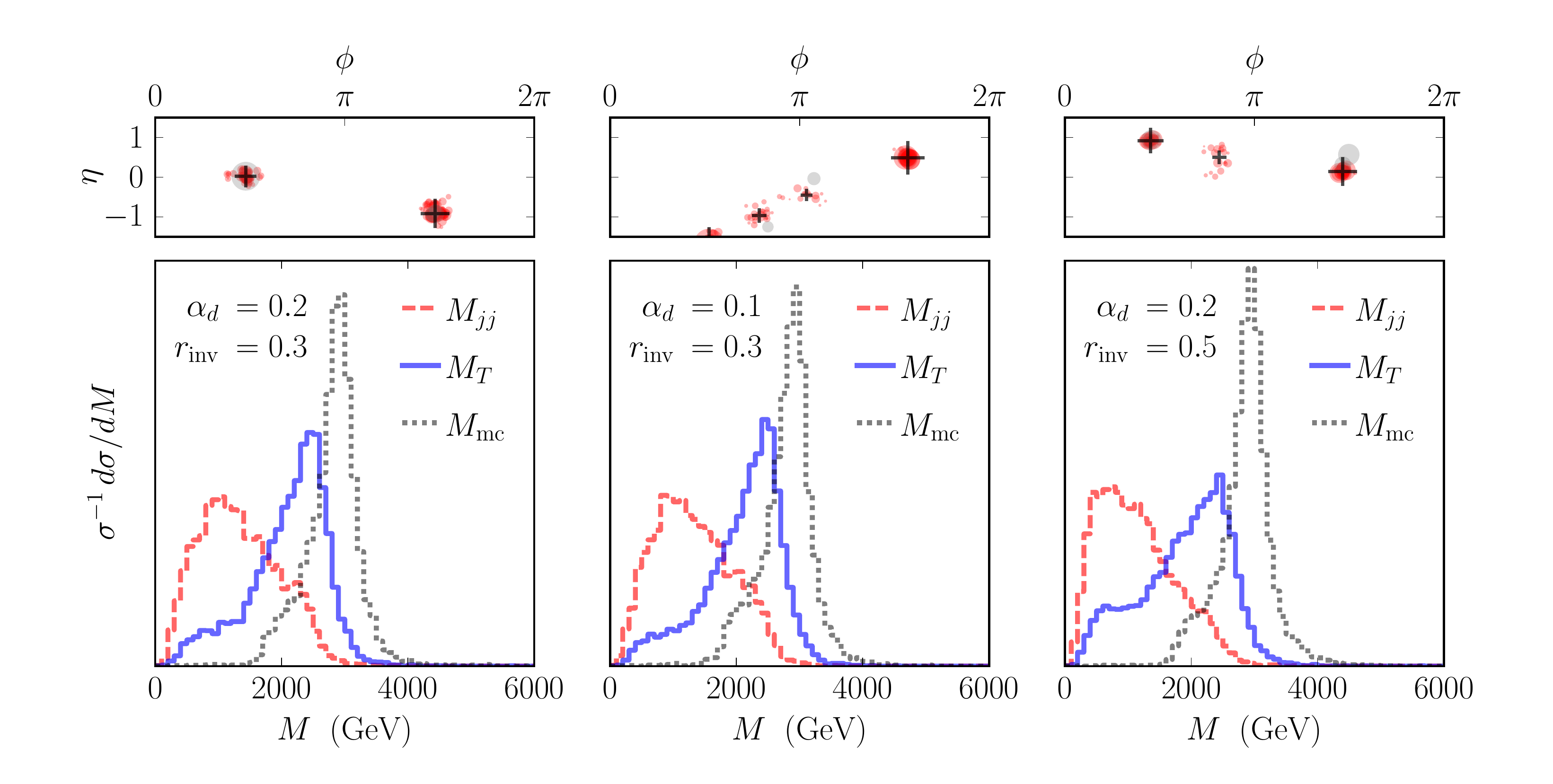}
\caption{Mass distributions after event selection cuts for the benchmark model in Table~\ref{table:param}, for various $\alpha_d$ and $r_\text{inv}$.  $M_{jj}$ is the mass of the two large reclustered jets, $M_{T}$ the transverse mass, and $M_\text{mc}$ the reconstructed $Z'$ mass using all the dark-matter particles in the Monte Carlo.  The $\eta -\phi$ lego plots show the corresponding energy deposition in the detector.  Red circles indicate visible SM hadrons, while the grey circles indicate undetected stable mesons.  The crosses indicate the position of anti-$k_T$ $R=0.5$ jets.  The relative size of each circle and cross is set by the $\sqrt{p_T}$ of the object. }
\vspace{-0.15in}
\label{fig:signal}
\end{figure*}

The $SU(2)_\D$ confines at a scale $\Lambda_\D \ll M_{Z'}$.  A QCD-like dark shower occurs when $M_i^2 \sim \Lambda_\D^2$ so that many dark gluons and quarks are produced, which subsequently hadronize.  Some of these dark hadrons are stable, while others decay back to the SM via an off-shell $Z'$.  The detailed spectrum of the dark hadrons depends on non-perturbative physics. Nonetheless, some properties of the low-energy states can be inferred from symmetry arguments.  There are two accidental symmetries: a dark-isospin number $U(1)_{1-2}$ and a dark-baryon number $U(1)_{1+2}$, where ``1" and ``2" refer to the $\chi_i$ flavor index.  For example, the mesons $\chi_1^\dagger \chi_1$ and $\chi_2^\dagger \chi_2$ are not charged under either of these symmetries, and are thus unstable.  The other mesons ($\chi_1 \chi_2^\dagger$, $\chi_1^\dagger \chi_2$) and baryons ($\chi_1 \chi_2$, $\chi_1^\dagger \chi_2^\dagger$) are charged under $U(1)_{1-2}$ and $U(1)_{1+2}$, respectively, and are stable.  

The spin of the dark mesons is also important.  Similar to the $\eta_b$~$(\eta_c)$ and $\Upsilon$~($J/\psi$) of the bottom (charm) system, the pseudoscalar and vector mesons should be degenerate.  By naive degree-of-freedom counting, the vector mesons are produced roughly three times as often as the pseudoscalar mesons~\cite{Carloni:2011kk}.  This impacts the phenomenology as the pseudoscalar decay is suppressed by a mass insertion, unlike the vector case.  Therefore, the pseudoscalar decays are dominated by $b$-quarks and are, in general, more displaced than the vector decays. The search strategy discussed below does not rely on $b$-tags or displaced tracks; it may be possible to take advantage of the pseudoscalar meson decays with a more sophisticated analysis.

Despite the myriad of possibilities for the dark sector, only certain parameters have a direct impact on the jet observables and missing transverse energy.  The strength of the dark shower, parametrized by $\alpha_\D$, plays a  critical role.  The coupling $\alpha_\D$ controls how many dark hadrons are emitted in the shower as well as their $p_T$ distributions, which has a direct and measurable impact on the jet observables.  In addition, the mass scale of the dark quarks affects the jet masses.

The number of dark-matter particles produced in the shower impacts $\vME_{T}$.  This effect can be parametrized as
\be
\label{eq:rinv}
r_\text{inv} \equiv \left\langle \frac{\text{\# of stable hadrons}}{\text{\# of hadrons} }\right\rangle.
\ee
The value of $r_\text{inv}$ depends on the details of the dark-sector model.  For the model described above with $M_1^2 = M_2^2$ , the average proportion of the stable and unstable hadrons is equal, implying $r_{\rm inv} \simeq 0.5$.  This assumes that the hadronization process is flavor-blind and that the dark quark masses are degenerate, and ignores baryon production, which is suppressed by a factor of $1/N_c^2$, where $N_c$ is the number of dark colors.
\begin{table}[b!]
\centering
\small
\renewcommand{\arraystretch}{2}
\setlength{\tabcolsep}{.2cm}
\begin{tabular}{c|c|c}
 & \textsc{description} & \textsc{benchmark\!}\\
\hline
$\sigma\times \text{Br}$ &
production rate  & 80 fb\\
\hline
$M_{Z'}$ & $Z'$ pole mass & 3 TeV\\
\hline
$M_{d}$ & dark hadron mass scale & 20 GeV\\
\hline
$\!\alpha_d(1\TeV)\!$ & running dark coupling & 0.2 \\
\hline
$r_{\rm inv}$ & fraction of stable hadrons & 0.3 
\end{tabular}
\caption{Parametrization for semi-visible jet search.}
\label{table:param}
\end{table}

A mass splitting between the flavors can lead to variations in $r_{\rm inv}$. Assuming $M_{2} \ge M_{1}$, in the Lund string model~\cite{Andersson:1983ia}, fragmentation into heavier dark quark pairs is suppressed by the factor
\begin{align}
T =\exp\left(-\frac{4\pi |M_{2}^2 - M_{1}^2|}{\Lambda_\D^2}\right).
\end{align}
Because of the exponential dependence of the fragmentation process, $r_\text{inv}$ is very sensitive to small splittings of the mass parameters.  As a result, fewer stable mesons are produced when $M_{2}^2-M_{1}^2 > \Lambda^2_\D $.  This decreases the value of $r_\text{inv}$ below 0.5.  To increase $r_\text{inv}$ above 0.5, one can increase the number of flavors $N_f$, thereby enlarging the number of stable mesons by $N_f(N_f-1)$, while only increasing the number of unstable mesons by $N_f$.  Clearly, $r_{\rm inv}$ can take on any value between $(0,1)$.  

Table \ref{table:param} summarizes the five parameters that are most relevant for semi-visible jet observables.  Three are sensitive to the details of the dark sector: the running dark-sector gauge coupling $\alpha_d(1 \TeV)$, $r_{\rm inv}$, and the mass scale for the dark mesons $M_d$. Note that by only including one value of $M_d$, we are assuming that the LHC will be insensitive to the dark spectrum mass splittings, \emph{i.e.}, $M_{Z'} \gg \Lambda_d$.   Additionally, there are two portal parameters: the production rate $\sigma \times \text{Br}$ and the $Z'$ mass. 

To perform a detailed collider study, $u\bar{u},d\bar{d} \rightarrow Z'\rightarrow \chi^\dag \,\chi$ events were simulated for the 14 TeV LHC using  {\tt PYTHIA8}~\cite{Sjostrand:2007gs} with the default \texttt{CTEQ6} parton distribution functions.  The dark-sector shower was simulated using the Hidden Valley {\tt Pythia} module \cite{Carloni:2010tw, Carloni:2011kk}, modified to include the running of $\alpha_\D$ as was done for \cite{Schwaller:2015gea}. Each meson had a probability $r_{\rm inv}$ to be a dark-matter particle.  The possible decays of dark baryons/mesons into each other were neglected.
The resulting particles were processed through {\tt DELPHES3}, with the CMS settings \cite{deFavereau:2013fsa}.

Anti-$k_{T}$ $R=0.5$ jets~\cite{Cacciari:2008gp} were constructed and then reclustered into two large jets~\cite{Nachman:2014kla} using the Cambridge/Achen (CA)  algorithm~\cite{Bentvelsen:1998ug} with $R=1.1$.  In a standard resonance search, one would use the invariant mass $M_{jj}^2=(p_1+p_2)^2$, where $p_{1,2}$ are the momenta of the two final large jets $j_{1,2}$.  However, the $M_{jj}$ variable is not useful when there are a significant number of dark-matter particles in the shower.  A variable that incorporates the missing momentum is the transverse mass: 
\begin{align}
M_T^2 = M_{jj}^2 + 2\left(\sqrt{M_{jj}^2 + p_{Tjj}^2}\, \ME_{T} - \vec{p}_{Tjj}\cdot \vME_{T}\right).
\end{align}
In a detector with perfect resolution, $M_{jj}\le M_T \le M_{Z'}$.  Figure~\ref{fig:signal} shows the distribution of $M_{jj}, M_{T}$ and $M_{\rm {mc}}$ after event selection. $M_{\rm {mc}}$ is the reconstructed $M_{Z'}$ computed from all the reclustered jets and truth-level dark-matter four-vectors.  $M_{T}$ in general yields a narrower, more prominent peak closer to $M_{\rm mc}$.  The top panels of Fig.~\ref{fig:signal} show sample events for the different signals.  The dark-sector particle multiplicity decreases for smaller $\alpha_\D$.
As $r_{\rm inv}$ is increased, the signal degrades because more stable mesons are produced and more information is lost. 

To estimate the reach at the LHC, we simulated $60 \times 10^6$ QCD events, $5\times 10^6$ $W^{\pm}/Z + jj$ events, and $5\times 10^6$ $t\bar{t}$ events. All samples were binned in $H_{T}$ in order to increase statistics in the high-$M_T$ tails~\cite{Avetisyan:2013onh} using \texttt{Madgraph5}~\cite{Alwall:2014hca} at parton level and {\tt PYTHIA8} for the shower and hadronization.  The dominant background after event selection is QCD and $W^{\pm}/Z + jj$.  For the signal, $25000$ events were generated for each choice of $M_{Z'}$ in increments of $500$ GeV, using the benchmark parameters in Table~\ref{table:param}.  An 8 TeV sample was used to validate the QCD background and limit-setting procedure~\cite{Cowan:2010js} against the CMS dijet resonance search~\cite{CMS-PAS-EXO-12-059}.  The $\ME_T$ distribution was also validated~\cite{CMS-PAS-JME-12-002}.  

Each event was required to have at least two $R=0.5$ anti-$k_{T}$ jets with $p_{T} > 200$ GeV and $|\eta|<2.5$, as well as $\ME_{T}>100$ GeV.  These pre-selection cuts model the impact of the trigger.  Then, the following cut-flow was applied:
\begin{itemize}
\item Recluster jets into $R=1.1$ CA jets ($j_1$, $j_2$);
\item Require $|\eta_{j_1}-\eta_{j_{2}}| < 1.1$; 
\item Require $\Delta \phi < 1$, where $\Delta \phi$ is the minimum azimuthal angle between $\vME_{T}$ and $\vec{p}_{Tj_{1,2}}$;
\item Veto isolated $e^{\pm}/\mu^{\pm}$ with $p_{T}>20$ GeV, $|\eta|<2.4$;
\item Require $\ME_{T} / M_{T} > 0.15 $.
\end{itemize}
The $R=1.1$ jets capture the wider radiation pattern expected from dark-shower dynamics. The cut on the pseudo-rapidity difference removes $t$-channel QCD~\cite{CMS-PAS-EXO-12-059, Aad:2014aqa}.  The lepton veto and $\Delta \phi$ requirements suppress electroweak backgrounds.  Finally, the $\ME_{T}/ M_{T}$ cut effectively acts as a missing energy requirement; cutting on the dimensionless ratio avoids sculpting the $M_T$ distribution.
\begin{figure}[t!]
\centering
\vspace{0.5cm}
%width = 0.42
\includegraphics[width=0.42\textwidth, trim = 15mm 10mm 0mm 20mm ]{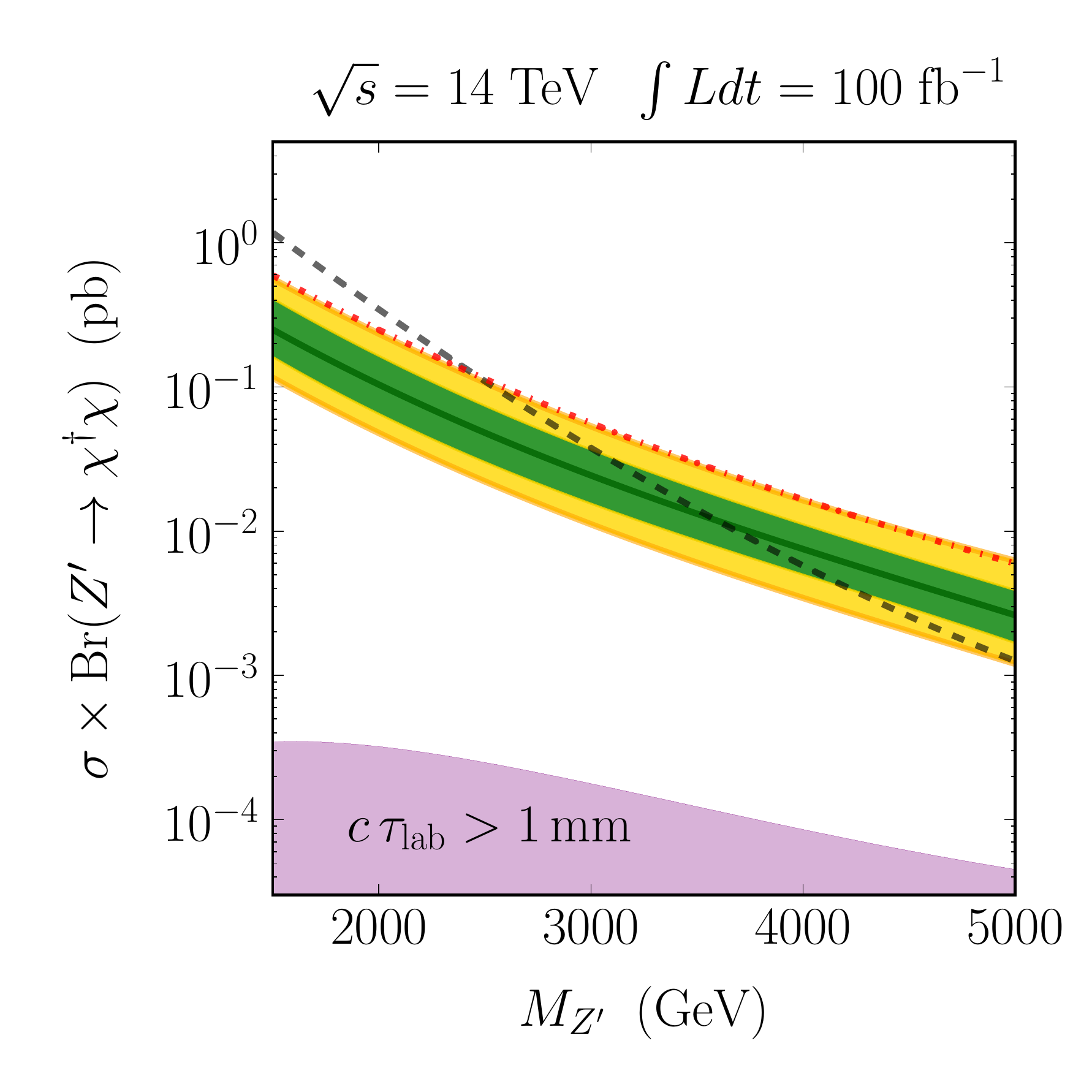}
\caption{Expected $\sigma\times \rm Br$ for the signal benchmark in Table~\ref{table:param}.  The shaded green (yellow) band corresponds to $\pm 1(2)$ standard deviations.  The dashed black line shows the $\sigma\times \rm Br$ for a $Z'$ with the same coupling to quarks as the SM $Z^0$; the dashed red line shows estimates $5\sigma$ discovery. The shaded purple region indicates where the vertices are displaced, assuming $g_{Z'}^\D\simeq1$ and that the shower is dominated by vector mesons.}
\label{fig:limits}
\vspace{-0.15in}
\end{figure}

After applying these cuts, a bump hunt was performed using $M_{T}$.  Following the dijet resonance searches ~\cite{Aaltonen:2008dn, CMS-PAS-EXO-12-059,Aad:2014aqa}, the resulting background distribution was parametrized using a fitting function---see Supplementary Material.
Assuming the background exactly follows the fit obtained from simulation, the exclusion reach for the signal benchmark can be computed.  Figure~\ref{fig:limits} shows the results for 100 fb$^{-1}$ of 14 TeV LHC data as a function of $M_{Z'}$ for the benchmark parameters (Table~\ref{table:param}).  We assume a $10\%$ width for the $Z'$, as computed using the benchmark parameters.  The production cross section times branching ratio for a $Z'$ with the same coupling as the SM $Z^0$ is shown as a reference.   A $Z'$ with SM couplings can be discovered (excluded) up to masses of $\sim 2.5 \TeV$ ($3.5 \TeV$).  

We estimate that the dijet limit on $\sigma\times\text{Br}(Z'\rightarrow q\bar{q})$ is comparable to the limit obtained for the dark-sector decay mode.  For $g^\D_{Z'} \simeq 1$, the branching ratio to the dark sector varies from 80\% to 50\% along the expected exclusion bound as the $Z'$ mass increases. Thus, the model would be discovered in the semi-visible jet channel \emph{before} it would be observed by the irreducible dijet channel; this conclusion only gets stronger for more integrated luminosity.

We simulate prompt decays for the dark mesons.  For a sufficiently heavy $Z'$ and small couplings, the dark vector meson decays could yield displaced vertices.  Requiring that the lab-frame decay length be $\lesssim\mathcal{O}$(1~mm), a lower bound on the couplings can be obtained:
\begin{align}
g^{{\SM}}_{Z'} \gtrsim
10^{-2}\left(\frac{1}{g^{{\D}}_{Z'}}\right)\sqrt{\frac{B}{10}}\left(\frac{M_{Z'}}{3 \textrm{ TeV}}\right)^2
\left( \frac{20 \textrm{ GeV}}{\Lambda_{\D}}\right)^\frac{5}{2}\!,
\label{eq:long_lived}
\end{align}
where $B\sim 10$ is the average boost factor computed from the benchmark simulation.  \eref{eq:long_lived} gives the lower purple region in Fig.~\ref{fig:limits}.  Modifications of the search strategy can still be effective in this region.

This Letter proposed a new search strategy for the discovery of hidden-sector physics in resonance searches.  In particular, the focus was on dark-sector showers that result in novel semi-visible jets---objects that are composed of SM hadrons and dark matter.  We argued that this generic signature could arise from a large class of strongly coupled dark-matter models.  Furthermore, we gave a simplified parameterization that allows for a systematic treatment of the signature space.  Finally, we provided expected exclusion limits using a bump hunt in transverse mass.  A $Z'$ with SM-size couplings to quarks could be discovered (excluded) up to $\sim$ 2.5 TeV (3.5 TeV).

There are two main extensions that can be explored.  First, one can allow for leptons, photons, and/or heavy-flavor particles to be produced in the shower.  Second, one can consider other production modes.  In this case, the semi-visible jets may not be aligned with the $\ME_T$ and additional variables using jet substructure, along the lines of~\cite{Fan:2011jc}, displaced vertices, and/or the presence of low-mass resonances may be necessary.   

With the LHC Run II on the horizon, it is important to rethink the program of dark-matter searches to guarantee that a wide range of new-physics scenarios are covered.  Non-trivial dynamics in the dark-matter sector is one of the many fantastic and unexpected ways that new physics can emerge.  This Letter provides a simple approach in preparation for this possibility.
\vspace{-0.26in}
\section*{Acknowledgments}
\vspace{-0.1in}
We are grateful to Pedro Schwaller, Daniel Stolarski, and Andreas Weiler for providing us with the code to run the dark-sector coupling as implemented for \cite{Schwaller:2015gea}.  We also thank Andrew Larkoski, Matthew Low, Duff Neill, Jesse Thaler, Scott Thomas, Tien-Tien Yu, and Daniel Whiteson for useful discussions.  TC is supported by an LHC Theory Initiative Postdoctoral Fellowship, under the National Science Foundation grant PHY-0969510. 

\vspace{0in}
\onecolumngrid
\vspace{0.3in}
\twocolumngrid
\def\bibsection{} 
\bibliographystyle{utphys}
\bibliography{SemiVisibleJets}

\onecolumngrid

\section*{\large Supplemental Material}
\vspace*{5pt}
\begin{spacing}{1.25}

\setcounter{equation}{0}
\setcounter{figure}{0}
\setcounter{table}{0}
\setcounter{section}{0}
\setcounter{page}{1}
\makeatletter
\renewcommand{\theequation}{S\arabic{equation}}
\renewcommand{\thefigure}{S\arabic{figure}}

\subsection{Razor Analysis}
Inclusive shape analyses can provide an alternative approach in the hunt for semi-visible jets.  One such example is the use of ``razor" variables, which were originally designed to extract signals such as pair production of squarks that decay into jets and neutralinos~\cite{Rogan:2010kb}.  The starting point is to recluster an event into two ``mega-jets," where the sum of the squared-invariant-mass of the jets is minimized~\cite{RazorWebsite}. Given these two mega-jets $j_1, j_2$, one can define the $M_R$ variable:
\begin{align}
M_R = \sqrt{\big(
|\vec p_{j_1}| + |\vec p_{j_2}|
\big)^2
- \big(
p_{z,j_1} + p_{z,j_2}
\big)^2
}.
\end{align}
For general new-physics signals, $M_R$ approximates the hard scale of the interaction. Additional discrimination between signal and background can be gained by considering the missing momentum $\vME_{T}$. For razor searches, the $R^2$ variable
\begin{align}
R^2 = \frac{1}{2\,M_R^2}
\left[ 
\ME_T \big(
p_{T,j_1} + p_{T,j_2}
\big)
 -
\vME_T \cdot \big(
\vec p_{T,j_1} + \vec p_{T,j_2}
\big)
\right]
\end{align}
is utilized.
$R^2$ characterizes the separation between $\vME_T$ and jet momenta. %SM backgrounds are concentrated near $(M_R ,R^2) \simeq (0, 0)$ and fall off exponentially for large $(M_R, R^2)$.  

\begin{figure}[b]
\centering
%width = 1.0
\includegraphics[width=0.45\textwidth, trim = 10mm 10mm 13mm 10mm]{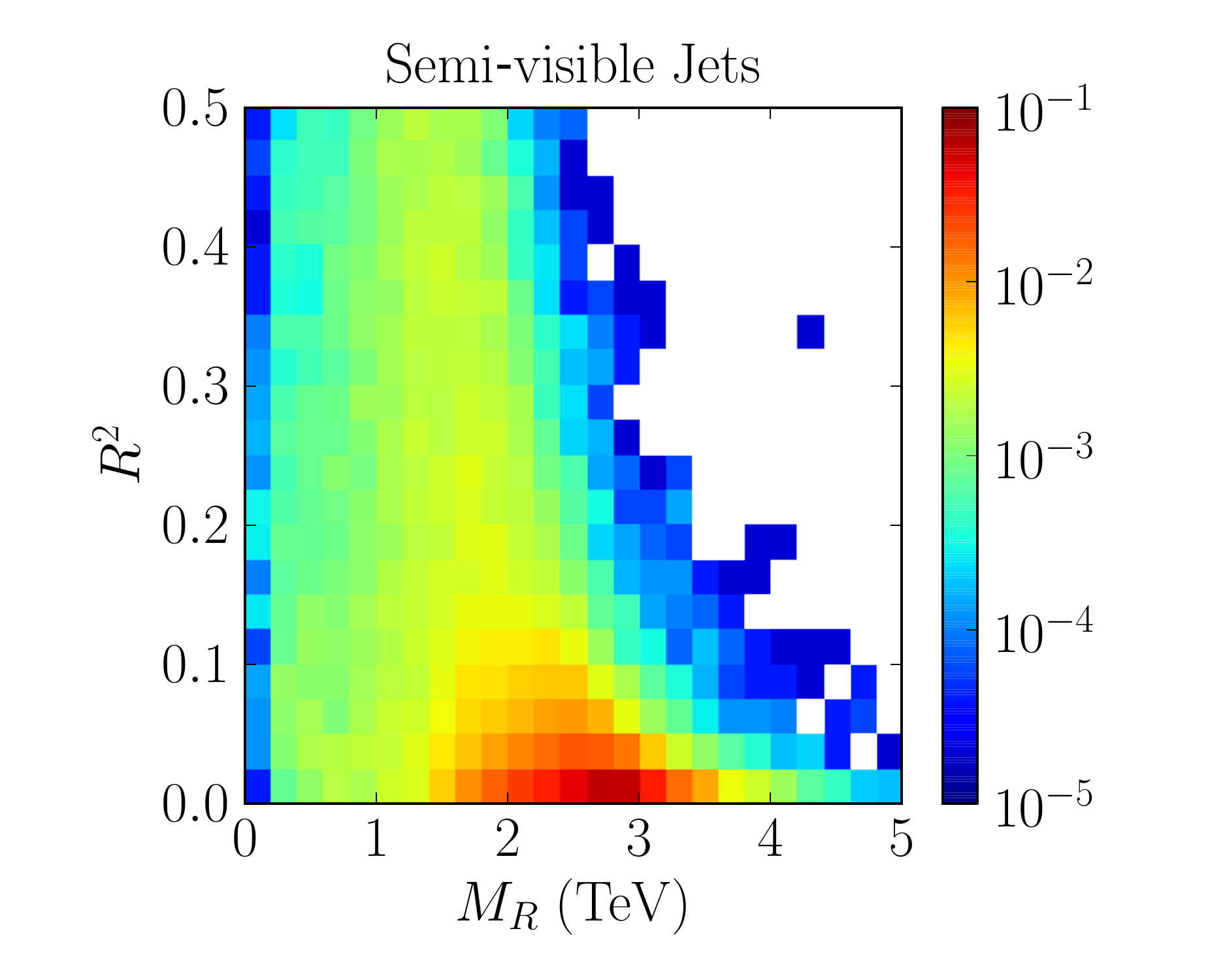}
\includegraphics[width=0.45\textwidth, trim = 10mm 10mm 13mm 10mm]{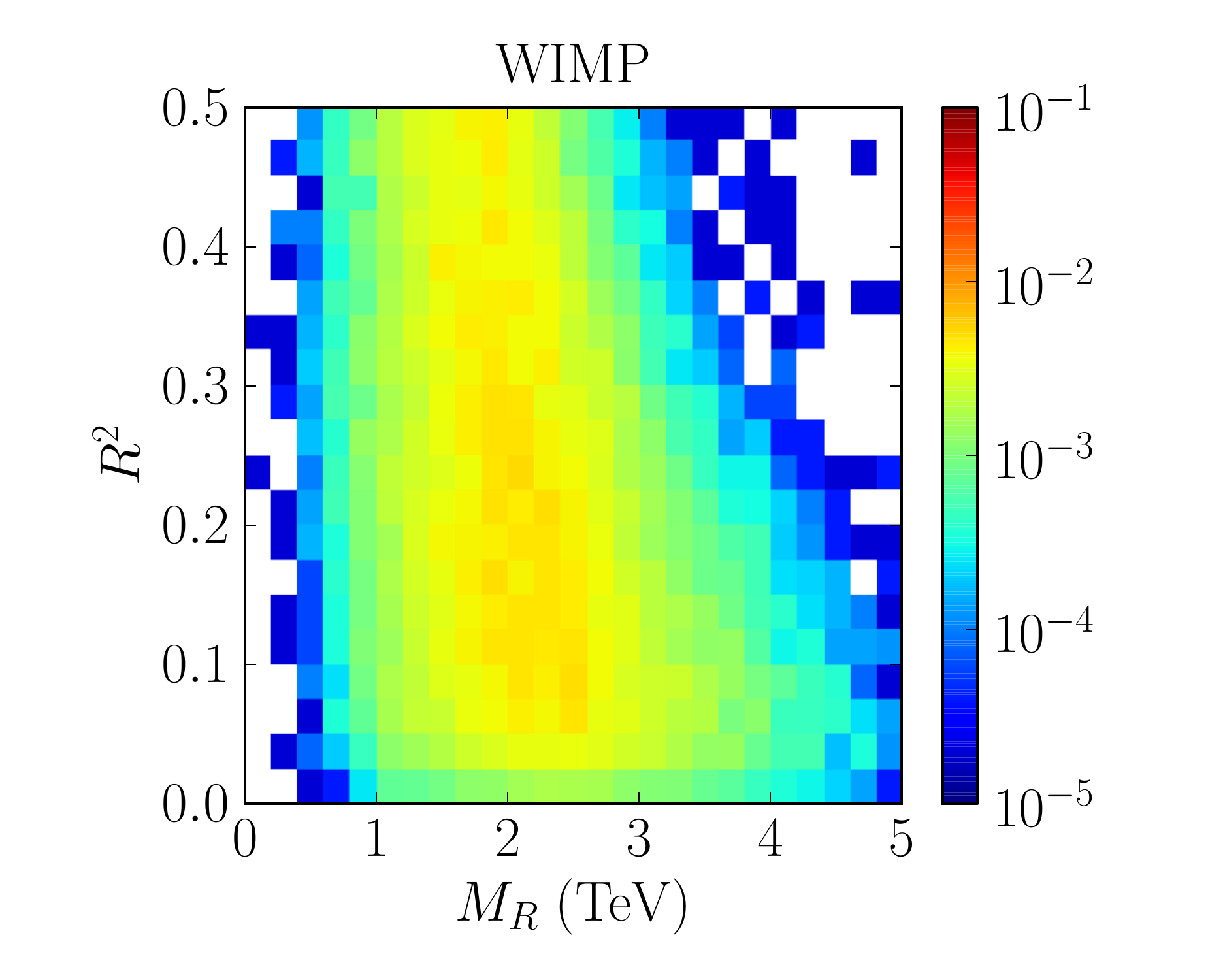}
\caption{The $R^2$ vs. $M_R$ distribution for the strongly coupled \emph{(left)} and weakly coupled \emph{(right)} dark-matter models from Figure~\ref{fig:supp_bkg} in the main Letter.  %Typically, the SM backgrounds (not shown) are concentrated near the origin, see \emph{e.g.},~\cite{Chatrchyan:2014goa}.  
The color scale denotes the fraction of events in each bin.}
\label{fig:supp_razor}
\vspace{-0.15in}
\end{figure}

Figure~\ref{fig:supp_razor} shows the distribution of $(M_R, R^2)$ for the strongly coupled (labeled semi-visible jet) and weakly coupled (labeled WIMP) models in the main Letter (see Fig.~\ref{fig:dphi}). The WIMP signal is broadly distributed in the bulk of the $(M_R, R^2)$ plane.  On the other hand, semi-visible jets yield a distinctive edge at $(M_R,R^2)=(M_{Z'}, 0)$,  with a tail extending to $R^2 > 0$ and $M_R < M_{Z'}$.  Current razor analyses delineate exclusive signal regions in the $(M_R, R^2)$ plane. For example, in the CMS dark-matter search that uses razor variables~\cite{CMS:2015fla}, the signal region is defined to have $R^2 > 0.5$ and $M_R \in (200, 300], (300, 400], (400, 600]$ and $(600,\infty) $ GeV.  Other analyses require less stringent cuts on $R^2$; for example, the supersymmetry analysis requires $M_R > 400$~GeV and $0.18<R^2<0.5$ for fully hadronic events~\cite{Chatrchyan:2014goa}.  None of these signal regions is ideal for the semi-visible jet signal, as they do not take advantage of the sharp kinematic feature in the $M_R$ distribution.  Re-optimizing the razor analyses may prove useful in distinguishing semi-visible jets from background; we leave this study for future work.

\subsection{Background Distributions}

\begin{figure}[h!]
\centering
\vspace{5mm}
%width = 1.0
\includegraphics[width=0.4\textwidth, trim = 1cm 10mm 0mm 10mm]{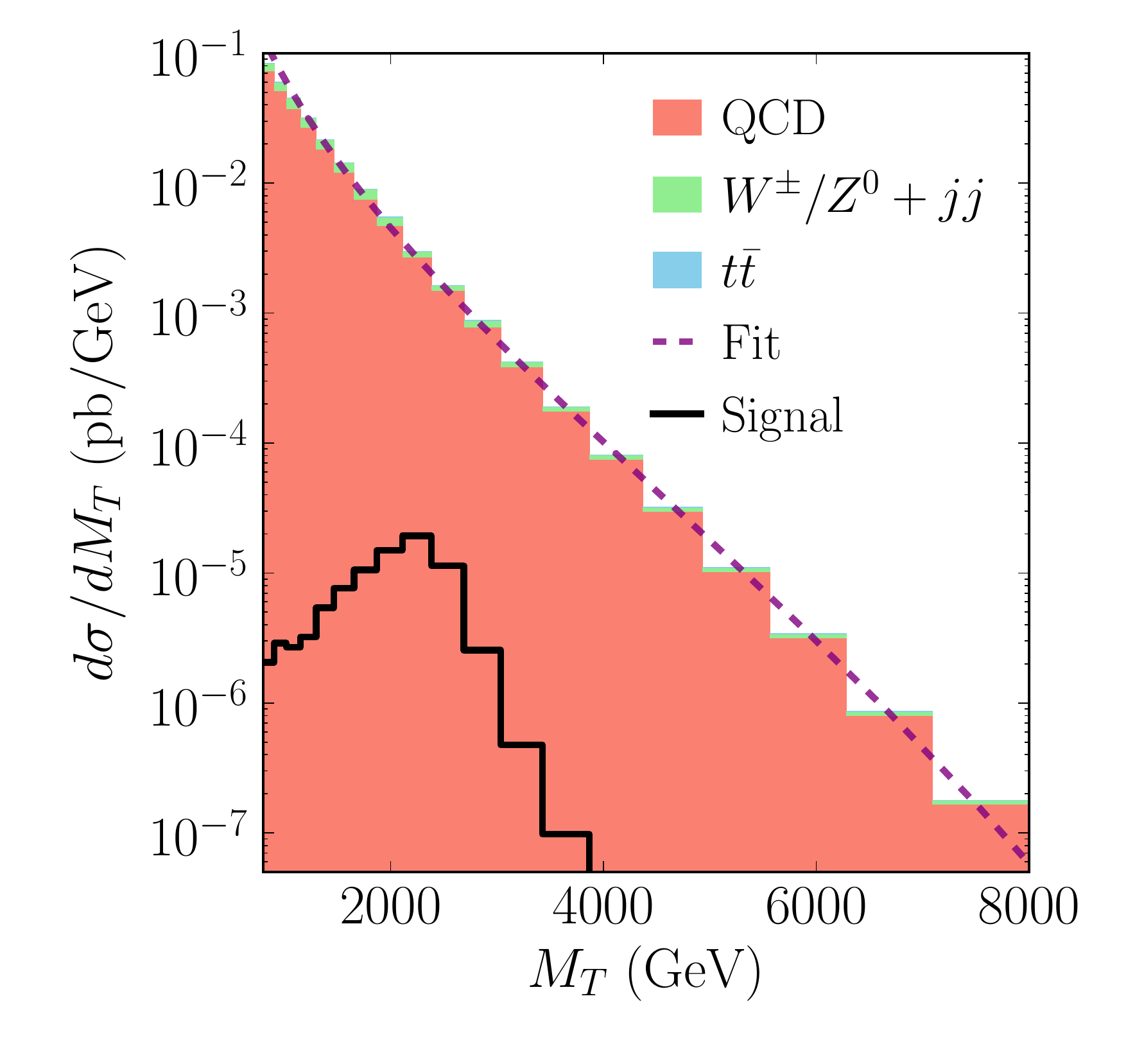}
\includegraphics[width=0.4\textwidth, trim = 1cm 10mm 0mm 10mm]{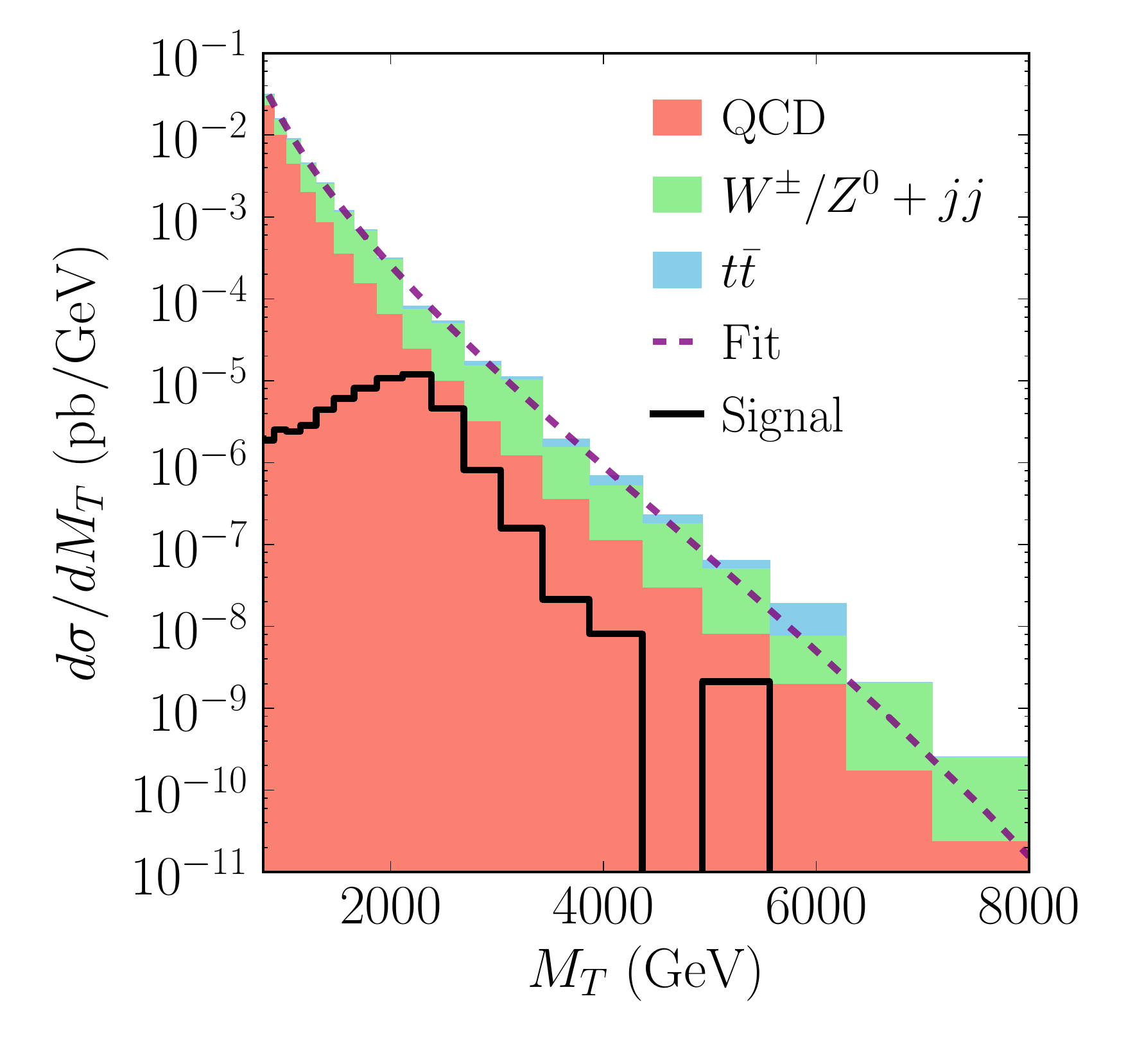}
\caption{$M_T$ distribution before \emph{(left)} and after \emph{(right)} an $\ME_{T}/M_T > 0.15$ requirement. (All other selection criteria are applied for both panels.)   The QCD, $W^\pm /Z^0 + jj$, and $t\bar{t}$ backgrounds are shown in red, green, and blue, respectively, and are stacked.  The dashed purple curve is an analytic fit to the total background using Eq.~\ref{eq:app_fit}. The solid black curve corresponds to the benchmark signal, with parameters listed in Table~\ref{table:param}.}
\label{fig:supp_bkg}
\end{figure}

Figure~\ref{fig:supp_bkg} shows the $M_T$ distributions of the backgrounds before and after the final $\ME_T/M_T$ cut.  Requiring $\ME_T/M_T>0.15$ removes a significant fraction of the QCD background; afterwards, the $W^\pm /Z^0 + jj$ and $t\bar{t}$ backgrounds dominate at large $M_T$.  The background distribution is modeled using the fitting function
\begin{equation}
f(x) = p_0 \frac{(1-x)^{p_1 + p_2 \ln x}}{x^{p_3 + p_4 \ln x}},
\qquad x=\frac{M_T}{\sqrt{s}}\, ,
\label{eq:app_fit}
\end{equation}
where the $p_i$ are fit parameters; the best-fit curve is shown in dashed purple.  The background distribution is smooth and the fit function provides a reasonable description in the region of interest. The $M_T$ distribution for the benchmark signal listed in Table~\ref{table:param} is shown in solid black.  The signal exhibits a prominent peak over the background and the $\ME_T/M_T$ cut significantly enhances the signal to background ratio.

\end{spacing}

\end{document}